\documentclass[preprint,aps]{revtex4}

\usepackage{graphicx}
\usepackage{dcolumn}
\usepackage{bm}
\usepackage{amsmath}
\usepackage{color}
\usepackage{slashbox}
\usepackage{multirow}

\begin{document}

\preprint{}
\title{Theory of edge states in a quantum anomalous Hall insulator/ 
spin-singlet $s$-wave superconductor hybrid system}
\author{Akihiro Ii$^{1}$, Keiji Yada$^{2}$, Masatoshi Sato$^{3}$, and Yukio Tanaka$^{1}$}
\affiliation{
$^1$ Department of Applied Physics, Nagoya University,Nagoya, 464-8603, Japan \\
$^2$ Venture Business Laboratory, Nagoya University,Nagoya, 464-8603, Japan \\
$^3$ Institute for Solid State Physics, University of Tokyo, Kashiwanoha 5-1-5, Kashiwa, Chiba 277-8581, Japan \\
}
\date{\today}

\begin{abstract} 
We study the edge states for a quantum anomalous Hall system (QAHS) coupled 
with a spin-singlet $s$-wave superconductor through the proximity effect, and 
clarify the topological nature of them. 
When we consider a superconducting pair 
potential induced in the QAHS, there appear  topological phases with 
nonzero Chern numbers, $i.e.$,  ${\cal N}=1$ and ${\cal N}=2$,
where Andreev bound states appear as chiral Majorana edge modes.  
We calculate the energy spectrum of the edge modes 
and the resulting local density of states. 
It is found that the degenerate chiral Majorana edge modes for ${\cal N}=2$ 
are lifted off by applying Zeeman magnetic field 
along the parallel to the interface or 
the shift of the chemical potential by doping. 
The degeneracy of the chiral Majorana edge modes and its lifting are
 explained by two different winding numbers defined at the
 time-reversal invariant point of the edge momentum.
%
%

\end{abstract}

\maketitle



%

%



\section{Introduction}
Andreev bound state (ABS) has been 
one of the central issue in 
superconductivity and condensed matter physics.  
It has been established that ABS is generated at the 
edge of unconventional superconductor where 
pair potential changes sign on the 
Fermi surface \cite{ABS,ABSR1,ABSR2,Hu,TK95}.  
ABS realized in spin-triplet chiral 
$p$-wave superconductor has remarkable properties 
since the ABS has a linear dispersion around $k_{y}=0$,  
where $k_{y}$ is a momentum parallel to the surface 
\cite{Matsumoto,Furusaki}. 
The time reversal symmetry 
is broken and spontaneous charge current is induced 
along the edge \cite{Furusaki}. 
Nowadays, this ABS has been recognized as a 
chiral Majorana edge mode \cite{Qi}. 
It is an analogous state to 
the chiral edge mode of Quantum Hall system (QHS) \cite{Girvin}. 
In QHS and its analogous superconducting state, the bulk edge
correspondence has been discussed based on the Chern number which is one of the 
topological numbers in condensed matter physics \cite{Thouless}. \par
It has been found recently that gapped time-reversal invariant systems also can
support gapless edge states.
In HgTe/CdTe quantum well, 
helical edge modes is generated due to the strong spin-orbit coupling. 
This system is so called 
Quantum spin Hall system (QSHS) \cite{Kane,Hasan}. 
There are also analogous systems in the world of superconductivity. 
In non-centrosymmetric (NCS) superconductors\cite{NCS1,NCS2},   
where the spin-orbit coupling is important as in the 
case of QSHS, the presence of helical Majorana
edge modes has been predicted \cite{Qi,Sato2009,Tanaka}. 
The helical Majorana 
edge mode is a special ABSs which is
expressed by 
two counter moving chiral Majorana fermions \cite{Qi}.
Instead of charge current, spin current is spontaneously generated 
along the edge. 
Several new features of  spin transport
stemming from these helical Majorana edge modes 
have been predicted \cite{Sato2009,Tanaka,Iniotakis,Eschrig,Yip}.
Furthermore, it has been clarified there are several types of 
helical Majorana edge modes with  \cite{Sato2009,Tanaka,Iniotakis,Eschrig,Yip} 
and without 
dispersion in NCS superconductors \cite{Mizuno,Yada,Schnyder2,Schnyder3}.
As well as chiral $p$-wave superconductors, the gapless edge modes in
NCS superconductors can
be characterized by the bulk topological numbers.


In general, superconducting states with topologically protected edge
states are dubbed as topological superconductor \cite{Schnyder,Qi,Sato} and the 
classification of topological superconductors 
has been done \cite{Schnyder}.  
In particular, chiral Majorana edge modes
have been a hot issue in the 
context of topological quantum computing \cite{Computation}. 
It has been known that 
 chiral Majorana edge mode is generated in the 
chiral $p$-wave pair potential \cite{Majoranap}. 
However, chiral $p$-wave pair potential is easily 
destroyed by the impurity scattering 
as compared to $s$-wave one. 
Furthermore,  the transition temperature 
of chiral $p$ -wave superconductor Sr$_{2}$RuO$_{4}$ \cite{Maeno} 
is rather low. 
Thus, chiral Majorana edge mode generated from simple 
spin-singlet $s$-wave superconductors is highly desired 
\cite{Majoranas,Majoranas1}. 
It has been proposed that a chiral Majorana edge mode is produced 
at the interface of  ferromagnet/spin-singlet $s$-wave superconductor 
junction on the substrate of three-dimensional topological insulator (TI) 
\cite{Majoranas}. 
Also, a simpler scheme using the Rashba spin-orbit interaction and the
Zeeman field has been proposed \cite{Majoranas1}. 
The essential point is the simultaneous presence 
of the broken inversion symmetry by the strong spin-orbit coupling 
and the time reversal symmetry breaking by ferromagnet or the Zeeman field. 

There is another way to realize chiral 
Majorana edge modes by using the chiral edge state of QHS attached with
spin-singlet $s$-wave pair potential.
Since the external magnetic field is very strong,  
it is difficult to induce pair potential for an ordinary QHS. 
However, as Qi {\it et.al.} has proposed \cite{Qi2}, this difficulty is 
overcome by 
considering a quantum anomalous Hall system (QAHS), instead of QHS,
where the exchange field is not so strong.
QAHS can be realized by the doping of magnetic impurity in 
QSHS \cite{Wu,Liu}. 
In this scheme, the presence of the 
chiral Majorana edge modes can be controlled by 
the band mass term $m$, chemical potential $\mu$ and the pair potential
$\Delta$.
The number of chiral Majorana edge modes can be 
classified by using the Chern number ${\cal N}$ \cite{Qi}. 
%
\begin{figure}[htbp]
\begin{center}
	\includegraphics[width=8.0cm]{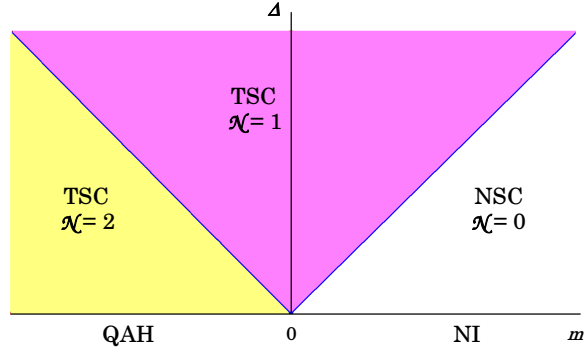}
	\caption{(Color online) Schematic illustration of 
the phase diagram of the QAH+SC hybrid system for $\mu$=0, proposed by Qi, Hughes and Zhang \cite{Qi}. 
The $x$ axis denotes the mass term $m$ and the $y$ axis denotes 
the magnitude of the superconducting pair potential $\Delta$. 
${\cal N}=2$ and ${\cal N}=1$ phases with $\Delta \neq 0$ 
are topological superconductor (TSC). 
${\cal N}=0$ phase is non topological superconductor (NSC). 
If $\Delta$=0, the present system changes from QAH state to normal insulator (NI).  For nonzero $\Delta$, the energy gap of the bulk closes 
at two boundaries of the phase diagram.  }
	\label{fig.1}
\end{center}
\end{figure}
\par

%
%
Althogh Qi $et.$ $al$ have proposed an interesting system, 
the physical properties of Majorana edge states and 
their relevance to observables have not been fully clarified yet. 
For this purpose, calculations of local density of states are 
important since it can be measured by angle resolved photoemission 
spectroscopy or scanning tunneling spectroscopy. 
In the following, we calculate the energy spectrum of the edge states 
and the resulting local density of states for various 
Chern number ${\cal N}$ 
in a QAHS coupled  with a spin-singlet $s$-wave superconductor. 
To clarify the difference between the 
${\cal N}=1$ and ${\cal N}=2$ states, we apply Zeeman magnetic field.
We find that when the direction of the 
magnetic field is parallel to the interface, the degeneracy 
of the two chiral Majorana edge modes in ${\cal N}=2$ states 
are  lifted off.  In order to understand the topological nature 
in detail, we evaluate the winding number of the bulk Hamiltonian. 
This number corresponds to the number of the zero energy state 
due to the presence of the bulk edge correspondence. 
We also clarify that the degeneracy of chiral Majorana 
edge modes in ${\cal N}=2$ can be lifted off by shift of chemical potential 
from zero.

\par
The organization of the paper is as follows. 
In Sec. \ref{sec:2}, we  introduce the model of QAHS with spin-singlet $s$-wave 
superconductor. In Sec. \ref{sec:3a} , we calculate the energy dispersion of the 
edge state and the corresponding local density of state at $\mu=0$ 
with and without Zeeman magnetic field. 
In Sec. \ref{sec:3b}, we introduce the winding number at 
$k_{y}=0$ in order to 
study the topological property of above edge states. 
In Sec. \ref{sec:3c}, we calculate the energy dispersion of the 
edge state for $\mu \neq 0$. 
In Sec. \ref{sec:4}, we summarize the results.  \par

\section{Formulation}
\label{sec:2}
We take a simple QAHS in the two-dimensional 
square lattice model, which is obtained by the replacement 
$p_{x,y} \rightarrow \sin p_{x,y} $ and $p^2_x+p^2_y \rightarrow
4-2(\cos p_x+\cos p_y)$ in the model used in \cite{Qi2}. 
Near the $\Gamma$ point, these replacements do not change 
the low energy physical properties of the system. 
Compared to continuum model, square lattice model is 
convenient when we calculate the local density of states.
The Hamiltonian describing low energy excitations of the quasiparticle is 
\begin{equation}
	{\cal H}_{QAH} = \sum_{{\bm p}} \psi^{\dagger}_{{\bm p}} h_{QAH}({\bm p}) \psi_{{\bm p}}\ ,
\end{equation}
\begin{eqnarray}
	h_{QAH} &=& {\bm d}({\bm p}) \cdot {\boldsymbol \sigma}  \nonumber \\
	&=& \begin{pmatrix}
		m({\bm p}) & A(\sin p_x-i\sin p_y) \\
		A(\sin p_x+i\sin p_y) & -m({\bm p}) \end{pmatrix}
\end{eqnarray}
with
\begin{equation}
	\psi_{{\bm p}} = \begin{pmatrix} c_{{\bm p}\uparrow} \\ c_{{\bm p}\downarrow} \end{pmatrix}
\end{equation}
where ${\bm d}({\bm p}) = (A\sin p_x, A\sin p_y,m({\bm p}))$, ${\boldsymbol
\sigma}$ is Pauli matrix, and $m({\bm p}) = m+B(4-2(\cos p_x+\cos p_y))$.
The band mass term $m({\bm p})$ determines
the magnitude of energy shift of up and down spin. 
$A$, $B$ and $m$ are material parameters.
The sign of $m$ changes
the topological property of the system.
Here note that the presence of $B$ term is crucial to create a QAHS. 
The energy dispersion of the above Hamiltonian is symmetric for the mass term $m$ for $B=0$, but is asymmetric for $B$$\neq$0. 
In other words, a nonzero value of $B$ makes the sign of $m$ meaningful.
We take $A$=$B$=1 in our following calculation.

In the following,  we consider the proximity effect by  
an attached spin-singlet $s$-wave superconductor, 
where pair potential is induced in the QAHS. 
The system is described by the BdG Hamiltonian,
\begin{equation}
	{\cal H}_{BdG} = \frac{1}{2} \sum_{{\bm p}} \Psi^{\dagger}_{{\bm p}} 
	\begin{pmatrix}
		h_{QAH}({\bm p})-\mu & \hat{\Delta} \\
		\hat{\Delta}^{\dagger} & -h^*_{QAH}(-{\bm p})+\mu \end{pmatrix}
	\Psi_{{\bm p}}
\end{equation}
with
\begin{equation}
	\hat{\Delta} = \begin{pmatrix}
		0 & \Delta \\
		-\Delta & 0 \end{pmatrix},
	\hspace{5mm} \Psi_{{\bm p}} = \begin{pmatrix} c_{{\bm
				      p}\uparrow} \\ c_{{\bm
				      p}\downarrow} \\
				      c^{\dagger}_{-{\bm p}\uparrow} \\
				      c^{\dagger}_{-{\bm p}\downarrow} \end{pmatrix}
\end{equation}
where $\mu$ is the chemical potential, $\Delta$ the induced pair potential 
of the spin-singlet $s$-wave superconductor. \par
\begin{figure}[htbp]
\begin{center}
	\includegraphics[width=8.5cm]{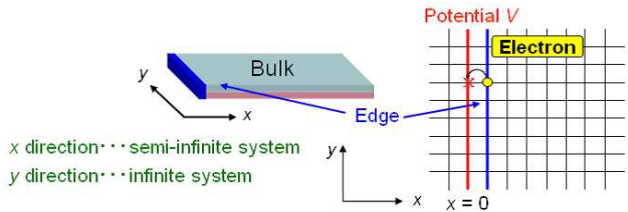}
	\caption{(Color online) Schematic illustration of QAHS with spin-singlet $s$-wave 
superconductor in 2D system. 
A red line denotes the infinite potential barrier 
and a blue line denotes the edge. 
In the actual calculation, 
the $x$-direction of the system is restricted finite width with 
$N_x$ ($N_{x}=4096$).  
We have checked that the  finite size effect is negligible. 
We use periodic boundary condition along the $y$-direction.}
	\label{fig.2}
\end{center}
\end{figure}

In order to calculate the local density of states (LDOS) at the 
edge, we introduce infinite potential along the $y$-axis as shown in 
Fig. \ref{fig.2}.  
We calculate the Green function at $x=1$ 
by $t$-matrix formulation \cite{Matsumoto1}.  
The system is infinite for the $y$-direction 
while it is semi-infinite for the $x$-direction (See Fig. \ref{fig.2}). 
Since translational invariance is absent along the $x$-direction,
only the momentum $k_y$ along the $y$-direction is a good quantum number. 
We express the Green functions by using spatial coordinate 
$x$ and $x'$ for fixed $k_y$ as follows; 
\begin{equation*}
	\hat{G}(x,x',k_y,\omega) = \hat{G}_0(x,x',k_y,\omega) \hspace{50mm}
\end{equation*}
\begin{equation}
	\hspace{15mm} -\hat{G}_0(x,0,k_y,\omega) \frac{1}{\hat{G}_0(0,0,k_y,\omega)} \hat{G}_0(0,x',k_y,\omega)
\label{eq:green1}
\end{equation}
with
\begin{equation}
	\hat{G}_0(x,x',k_y,\omega) = \frac{1}{N_x} \sum_{k_x} e^{ik_x(x-x')} \hat{G}_0(k_x,k_y,\omega)\ ,
\end{equation}
\begin{equation}
	\hat{G}_0(k_x,k_y,\omega) = \frac{1}{\omega-{\cal H}(k_x,k_y)}
\end{equation}
where $N_x$ is a total number of 
lattice points for the $x$-direction. In the right hand side of
Eq.(\ref{eq:green1}), 
the first term denotes the unperturbed bulk 
Green function, and the second term is
the scattering effect at the edge. 
The momentum resolved 
LDOS at the edge $x=1$ $N(k_{y},\omega)$  is written as
\begin{equation}
N(k_y,\omega) = -\frac{1}{\pi} \left.\left( {\rm Im} [ \hat{G}^R_{11}(x,x,k_y,\omega) ] + {\rm Im} [ \hat{G}^R_{22}(x,x,k_y,\omega) ] \right)\right|_{x=1}
\end{equation}
where the upper suffix $R$ means retarded Green function; 
replacing $\omega$ to $\omega + i\delta$ with infinitesimal positive number $\delta$, 
and the lower indices 11 and 22 indicate the matrix elements. 
And we obtain the LDOS at the edge for energy $\omega$ measured from 
the Fermi level as follows
\begin{equation}
	D(\omega) = \frac{1}{N_y} \sum_{k_y} N(k_y,\omega)
\end{equation}
where $N_y$ is a total number of lattice points for the $y$-direction. We set $N_x$=$N_y$=4096 in the actual calculations. 
In the next section, we will show the spectrum $N(k_{y},\omega)$ 
and $D(\omega)$. 
%
%
%
\section{Results}
\subsection{edge states for $\mu=0$}
\label{sec:3a}
%
%
The minimum value of the bulk energy gap $E_{g}$ 
is given at $kx=ky=0$ as far as we have studied. 
In this case $E_g = |m+\Delta|$ or $|m-\Delta|$. 
We fix $\Delta=0.25$ and $\mu=0$. 
We change the value of $m$ for 
(a) $m=-0.5$, (b) $m=0$, and (c) $m=0.5$ 
corresponding to ${\cal N}=2$, ${\cal N}=1$ and ${\cal N}=0$ phases,
respectively. 
In every case, $E_{g}$ is 0.25 \cite{Qi2}. 
The bright line between upper and lower bands 
corresponds to the chiral Majorana edge modes 
for $-0.25<\omega<0.25$. 
On the other hand, the background structure with 
parabolic shape of the spectrum with $\mid \omega \mid> 0.25$ 
expresses the continuum level originating from 
energy bands of the bulk QAHS.  
\begin{figure}[htbp]
\begin{center}
	\includegraphics[width=9.0cm]{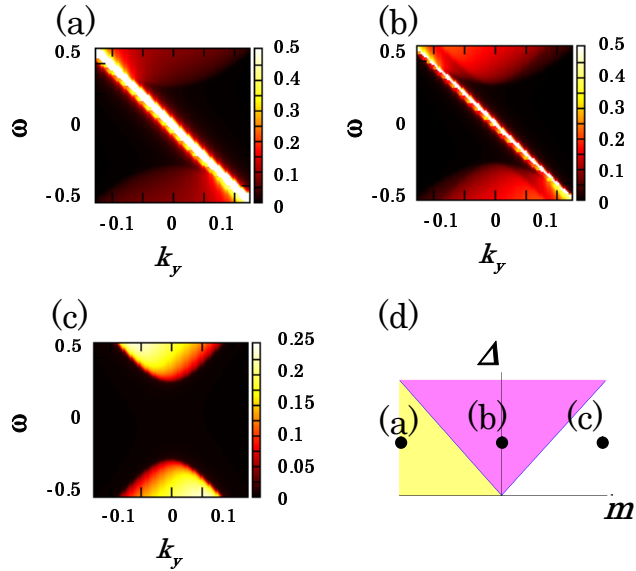}
	\caption{Momentum resolved LDOS at the edge $N(k_{y},\omega)$ 
is plotted as a function of $\omega$ and $k_{y}$  for $\Delta=0.25$. 
(a):$m$=$-$0.5, (b):$m$=0, and (c):$m$=0.5; these three cases are 
 shown on $m$-$\Delta$ space in (d), where (a), (b) and (c) 
belong to ${\cal N}=2$, ${\cal N}=1$ and ${\cal N}=0$ case,  respectively. 
(d)Schematic phase diagram for $\mu=0$. }
	\label{fig.3}
\end{center}
\end{figure}
\begin{figure}[htbp]
\begin{center}
	\includegraphics[width=9.0cm]{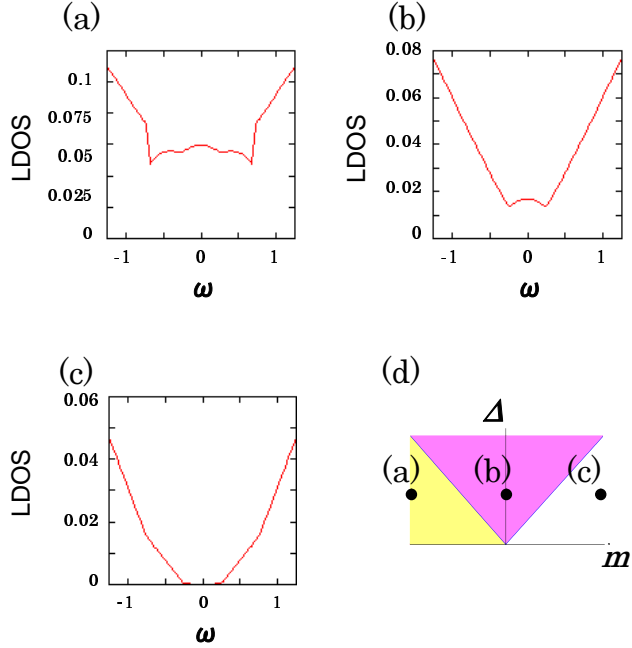}
	\caption{(Color online) LDOS at the edge. (a)$m=-0.5$, 
(b)$m=0$, and (c)$m=0.5$. 
At these parameters, the gap of the bulk bands is 0.25 in all cases. 
For (a) and (b), there is no energy gap in LDOS. 
}
	\label{fig.4}
\end{center}
\end{figure}
\noindent
We can see the chiral Majorana edge modes 
for (a) and (b), but no edge mode in (c) as expected by 
the Chern number \cite{Qi2}. 

It is interesting to look at  LDOS at the edge $D(\omega)$ 
since it can be observed by scanning tunneling spectroscopy (STS). 
In Fig. \ref{fig.4}, we plot $D(\omega)$ with the same parameters used
in Fig. \ref{fig.3}.
For $m=-0.5$ with ${\cal N}=2$, the resulting 
$D(\omega)$ has a finite value at $\omega=0$. 
It has a peak at $\omega=0$. 
Similar to this case, for $m=0$ with ${\cal N}=1$, 
LDOS has a peak structure at $\omega=0$. 
On the other hand, for $m=0.5$, 
LDOS has a gap structure 
where $D(\omega)=0$ for $\mid \omega \mid < 0.25$. 
The absence of peak structure at $\omega=0$ is consistent with 
${\cal N}=0$, where there is no chiral Majorana edge mode. 
The presence of the chiral Majorana edge mode seriously 
changes the resulting zero energy LDOS at the edge. 
Thus, it is possible to check the presence of chiral Majorana edge modes 
by STS. 
However, there is no qualitative difference between 
${\mathcal N}=2$ and ${\mathcal N}=1$ phases as seen from LDOS. 
From the discussion based on the Chern number, 
we can expect  that there are degenerate two edge states in 
${\mathcal N}=2$ phase. \par
To discriminate ${\cal N}=1$ phase from ${\cal N}=2$ phase, 
we need a further idea:
We apply the magnetic field to the system to make 
sure the difference between these two phases. 
We add the Zeeman term 
$\mu_{B} {\bm H} \cdot {\boldsymbol \sigma}$ 
in the original Hamiltonian, 
where $\mu_B$ is the Bohr magneton, ${\bm H}$ is the Zeeman magnetic field. 
\begin{figure}[htbp]
\begin{center}
	\includegraphics[width=9.0cm]{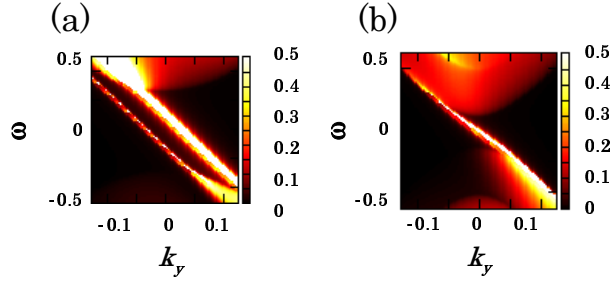}
	\caption{(Color online) Momentum resolved LDOS at the edge
 $N(k_{y},\omega)$ in the presence of magnetic field along the
 $y$-direction at $\Delta$=0.25 and $\mu_{B}|{\bm
 H}|=0.15$. (a):$m$=$-$0.5 and (b):$m$=0.}
\label{fig.5}
\end{center}
\end{figure}
The magnitude of the 
bulk energy gap $E_{g}$ is influenced by the Zeeman magnetic field. 
When ${\bm H}$ is along $y$-direction, 
$E_{g}$ is given by
$E_{g}=\mid \sqrt{m^{2}+(\mu_{B} |{\bm H}|)^{2}} - \Delta \mid$ 
In Fig.\ref{fig.5}, $N(k_{y},\omega)$ 
is plotted for $\Delta=0.25$ and $\mu_{B}|{\bm H}|=0.15$, where 
magnetic field is applied along the $y$ direction. 
The resulting $E_{g}$ is 0.27 and 0.1 for 
(a) $m=-0.5$ and (b) $m=0$, respectively. 
As seen from Fig. \ref{fig.5}(a), for ${\cal N}=2$ phase,
a splitting of the degenerate two chiral Majorana edge 
modes appears in the momentum resolved 
LDOS $N(k_{y},\omega)$.  
On the other hand, for ${\mathcal N}=1$ phase, 
single Majorana edge mode remains the same even in the presence of 
the magnetic field. 
Thus, we can discriminate ${\cal N}=1$ and ${\cal N}=2$ phase 
through the momentum resolved LDOS at the edge in the 
presence of Zeeman magnetic field. 
%
We also found that the splitting of Majorana edge modes for ${\cal N}=2$
phase appears only when the direction of the magnetic field is along the
$y$ direction.
In order to understand the orientational dependence of the 
magnetic field on the energy spectrum, 
we introduce the winding number of the system at $k_{y}=0$ 
in the following subsection.
\subsection{Winding number}
\label{sec:3b}
The winding number $W$ is one of the topological invariants 
that is well defined for the system with odd-number of spatial dimension 
\cite{Sato1}. 
It can count the number of zero energy state 
protected by topological property of the Hamiltonian. 
Although the dimension of our present system is two, if we 
fix $k_y$ to a certain value, we can express the present system as an 
effective one-dimensional Hamiltonian \cite{Sato1}. 
In order to define $W$, we look for a hermitian matrix $\Gamma$ 
which anti-commutes with the Hamiltonian; 
\begin{equation}
	\{ H({\bm k}),\Gamma \} = 0. 
\end{equation}
In the presence of the time reversal (TR) symmetry $\Theta$, 
and particle-hole symmetry $C$, we can choose $\Gamma$ as 
$\Gamma=-iC\Theta$. 
However, in the present system, the time reversal symmetry is broken by 
$m({\bm p})$. Thus, we must find other $\Gamma$s. 
Since we found such $\Gamma$s only for $k_y=0, \pi$, 
we focus on these cases in the following discussions. 
%
%
The searching of $\Gamma$ is done as follows.
Since $\Gamma$ is a $4\times 4$ hermitian matrix, it can
be expressed  by direct products of two Pauli matrices, 
\begin{equation}
	\Gamma_{\mu\nu} = \sigma_{\mu} \otimes \tau_{\nu}
\end{equation}
where $\sigma_{\mu}$ ($\mu=0,1,2,3$) operates on the spin space, and
$\tau_{\mu}$ ($\mu=0,1,2,3$) on the particle-hole  space.
%
The suffix 
0 indicates the unit matrix, and 1, 2 and 3 the $x$, $y$, and $z$
components of the Pauli matrices, respectively.
%
%
In a similar manner, the Hamiltonian can be expressed by the basis of
$\Gamma_{\mu\nu}$. 
Then from the (anti-)commutation relations between the Pauli matrices,
we find that 
only $\Gamma_{23}$ and $\Gamma_{32}$ anti-commute with the Hamiltonian. 
These matrices also anti-commute with the Hamiltonian even in the 
presence of Zeeman magnetic field along the $z$ direction since 
the applied Zeeman term just changes the mass term in the Hamiltonian.
On the other hand, if we apply the Zeeman 
magnetic field along the $x$-axis, 
only $\Gamma_{23}$  anti-commutes with the Hamiltonian. 
Furthermore, if the direction of the applied magnetic field 
is the $y$-direction, only $\Gamma_{32}$ anti-commutes.  
As is shown below, these differences imply that 
the topological nature of the Hamiltonian strongly depends on the 
direction of the Zeeman magnetic field. \par 

Using these $\Gamma$s, we can define the winding number for each $\Gamma$. 
First, we diagonalize these $\Gamma$ matrices;
\begin{equation}
	U_{\Gamma}^{\dagger} \Gamma U_{\Gamma} = 
	\begin{pmatrix}
		{\bm I}_{2 \times 2} & 0 \\
		0 & -{\bm I}_{2 \times 2}
	\end{pmatrix}
\end{equation}
and transform the Hamiltonian to anti-diagonalized form by $U_{\Gamma}$.
\begin{equation}
	U_{\Gamma}^{\dagger} H({\bm k}) U_{\Gamma} = 
	\begin{pmatrix}
		0 & q({\bm k}) \\
		q({\bm k})^{\dagger} & 0
	\end{pmatrix}.
\label{eq:UHU}
\end{equation}
Then we calculate the determinant of the sub matrix $q(\bm{k})$  in
Eq.(\ref{eq:UHU}),
and denote its real part as $m_1$ and imaginary part as $m_2$. 
\begin{equation}
	\det |q({\bm k})| \equiv m_1({\bm k}) + i m_2({\bm k})
\end{equation}
Finally, we obtain the winding number $W$ as the following integral;
\begin{equation}
	W = \frac{1}{2\pi} \int^{\pi}_{-\pi} \frac{\partial \theta({\bm k})}{\partial k_x} \Biggr|_{k_y=0} dk_x
\end{equation}
where
\begin{equation}
	\theta({\bm k}) \equiv \arg \det |q({\bm k})| = \tan^{-1}\frac{m_2({\bm k})}{m_1({\bm k})}.
\end{equation}
Here we only consider the winding number at $k_y=0$ since they are found
to be zero at $k_y=\pi$ in the following cases. 
We denote the winding number for 
$\Gamma_{23}$ and $\Gamma_{32}$ as $W_{23}$ and $W_{32}$, respectively. 
\begin{figure}[htbp]
	\begin{center}
	\includegraphics[width=8cm]{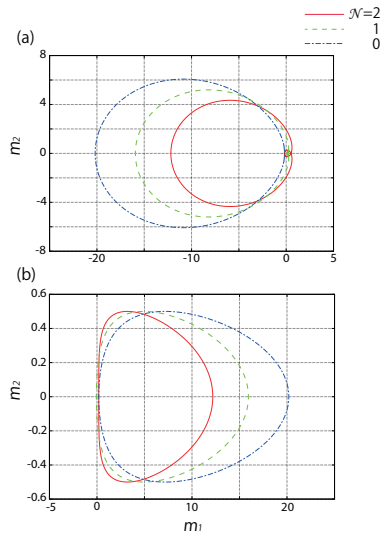}
	\caption{(Color online) Contour plots of  $m_1({\bm k})$ and $m_2({\bm k})$ for
	 (a) $W_{23}$ and (b) $W_{32}$. Here we fix $k_y$ as $k_y=0$ and
	 changes $k_x$ from $-\pi$ to $\pi$.}
	\label{fig.6}
	\end{center}
\end{figure}
The winding numbers $W_{23}$ and $W_{32}$ become visible 
by plotting $m_1$ and $m_2$ in the first Brillouin zone as shown in
Fig. \ref{fig.6}. 
To make it clear, we enlarge the scale of Fig. \ref{fig.6} around the
origin, in Fig.\ref{fig.7}.
The number of the rotation of the contour around the origin just
corresponds to the winding number.
\begin{figure}[htbp]
	\begin{center}
	\includegraphics[width=8cm]{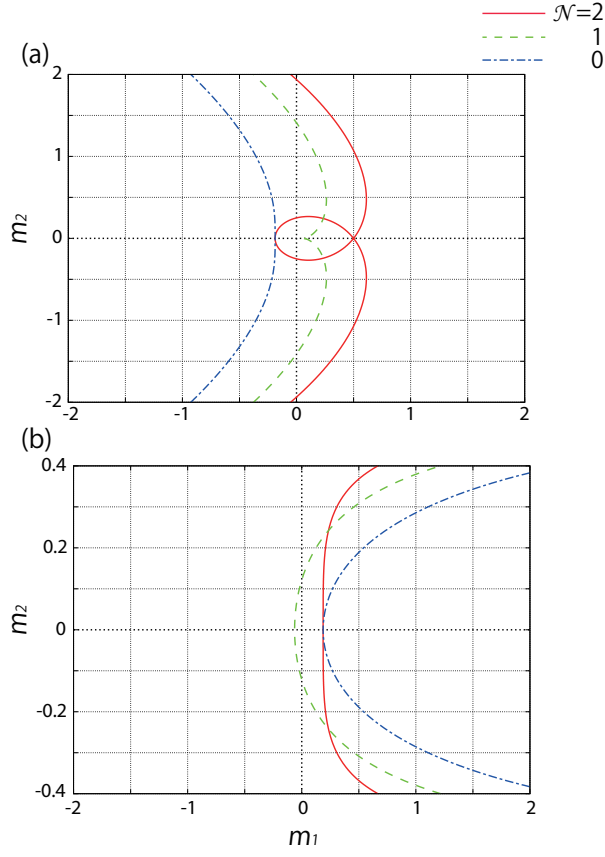}
	\caption{(Color online) Enlarged plots of Fig. \ref{fig.6}. }
	\label{fig.7}
	\end{center}
\end{figure}
The obtained $W_{23}$ and $W_{32}$ are summarized in  Table \ref{table1}.
\begin{table}
\begin{center} 
\begin{tabular}[t]{|c||c|c|c||c|c|c|}
\hline
 & ${\cal N}=2$ & ${\cal N}=1$ & ${\cal N}=0$ 
& ${\bm H}\parallel \hat{\bm x}$
& ${\bm H}\parallel \hat{\bm y}$
& ${\bm H}\parallel \hat{\bm z}$\\
\hline
$W_{23}$ &2 & 1 & 0 & $\bigcirc$ &$\times$ &$\bigcirc$ \\ 
\hline
$W_{32}$ &0 & 1 & 0 &$\times$ &$\bigcirc$ &$\bigcirc$ \\ 
\hline
\end{tabular} 
\end{center}
\caption{Correspondence between the winding number $W$ and the Chern
 number ${\mathcal N}$. We also show the validity of the winding numbers
 in the presence of the Zeeman magnetic field.
$W_{23}$ can not be defined for the applied magnetic field along $y$-direction, while $W_{32}$ can not be defined for the applied magnetic field along 
the $x$-direction. 
}
\label{table1}
\end{table}

%
From the winding numbers in Table \ref{table1}, we can derive  
the following results. 
Let us first consider the case without the Zeeman magnetic field.
In this case, both of $W_{23}$ and $W_{32}$ are well-defined.
Then, the bulk-edge correspondence tells us that 
the number of zero energy states at $k_y=0$ should be consistent with
these winding numbers.
As a result, there should be two zero energy edge states at $k_y=0$ for
${\cal N}=2$ phase, and one zero energy edge state for ${\cal N}=1$. 
Here note that $W_{32}=0$ for ${\cal N}=2$ does not necessarily mean no
zero energy edge states at $k_y=0$.
To be consistent with $W_{23}=2$ for ${\cal N}=2$ at the same time, we
need to have two
zero energy edge states at $k_y=0$.
These two zero modes excellently agree with the two-fold degeneracy of the
chiral Majorana edge modes found in Fig.\ref{fig.3}(a).  

Now consider the case with a weak Zeeman magnetic field.
As was mentioned above, $W_{23}$ becomes ill-defined if we apply the
Zeeman magnetic field in the $y$-direction.
Thus, the number of the zero energy state at $k_y=0$ is determined solely
by $W_{32}$ in this case. 
It is found that the remaining winding number $W_{32}$ takes
the same value as that without the Zeeman magnetic field, so 
for ${\cal N}=2$ phase, 
the zero energy states at $k_y=0$ should vanish. 
This result excellently agrees again with the lifting of the degeneracy
of the chiral Majorana modes in Fig.\ref{fig.5}(a). 
These winding numbers also explains why the degeneracy is
not lifted off if the Zeeman magnetic field is along the $x$ or
$z$-direction: In these cases, $W_{23}$ remains
well-defined, so the two degenerate zero modes also remain at $k_y=0$.

%
Finally, we discuss the edge state in  ${\mathcal N}=1$ phase. 
For any weak Zeeman magnetic field, at least one of
$W_{23}$ and $W_{32}$ is well-defined, and both of them take 1. 
Thus, 
a single zero energy edge mode is always
ensured at $k_y=0$.
In this sense, 
the edge state at $k_{y}=0$ in ${\mathcal N}$=1 phase is 'robust' against
perturbation. 


%
%
%
\subsection{edge states for finite $\mu$}
\label{sec:3c}
\begin{figure}[htbp]
\begin{center}
	\includegraphics[width=9.0cm]{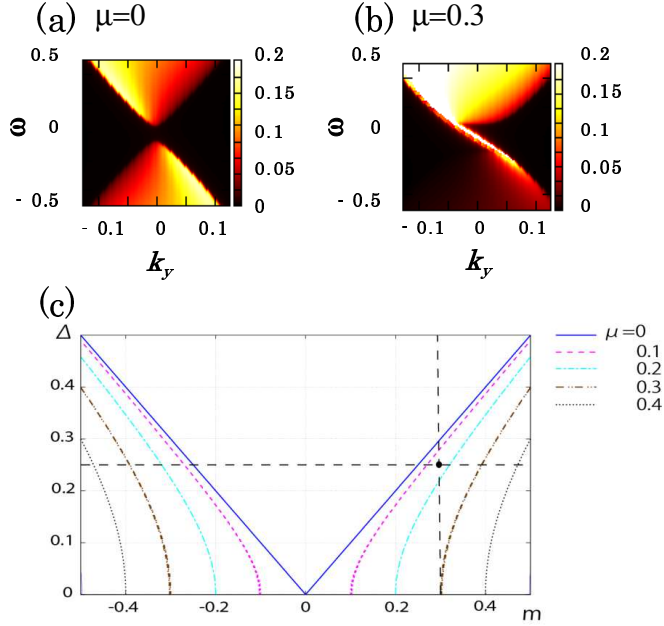}
	\caption{(Color online) Energy spectrum of the edge modes are 
plotted as a function of $k_{y}$ with $m$=0.3 and $\Delta$=0.25 
for (a)$\mu=0$, and (b)$\mu=0.3$. 
(c)phase boundary of ${\cal N}=2$, ${\cal N}=1$, and 
${\cal N}=0$ phases are 
plotted as a function of $\Delta$ and $m$ for various $\mu$ \cite{Qi2}.  
}
\label{fig.8}
\end{center}
\end{figure}

In this subsection, we consider $\mu \neq 0$ case, 
which corresponds to a doped QAHS case. 
It has been clarified that 
the condition for closing of the bulk band gap is as follows \cite{Qi2};\\%
\begin{equation}
\Delta^2+\mu^2=m^2 \hspace{10mm} {\rm when} \ \mu \neq 0.
\end{equation}
The momentum resolved LDOS $N(k_{y},\omega)$ 
is plotted in Figs. \ref{fig.8}(a) and (b)
for (a) $\mu=0$ and 
(b) $\mu=0.3$, respectively. (We set here $m=0.3$ and $\Delta=0.25$.) 
Since the magnitude of the bulk energy gap $E_{g}$ 
is given by 
the minimum value of 
$E_{g}=\mid m + \sqrt{ \mu^{2}+\Delta^{2}} \mid$
and $E_{g}=\mid m - \sqrt{ \mu^{2}+\Delta^{2}} \mid$, 
the resulting $E_{g}$ 
is 0.05 and 0.09 for (a) and (b), respectively. 
There is no chiral Majorana edge state in Fig. \ref{fig.8}(a), but  
it exists in Fig. \ref{fig.8}(b). 
Therefore, with the change of $\mu$, 
the chiral Majorana edge state is generated. 
In other words, the change of $\mu$ introduces the 
quantum phase transition of the 
topological number from ${\mathcal N}=0$ to ${\mathcal N}=1$.  
Actually, the boundaries of the three phases ($i.e.$ ${\cal N}=0,1,2$ ) depend on 
the values of $\mu$ in addition to $m$ and $\Delta$.
See Fig. \ref{fig.8}(c).

\begin{figure}[htbp]
\begin{center}
	\includegraphics[width=9.0cm]{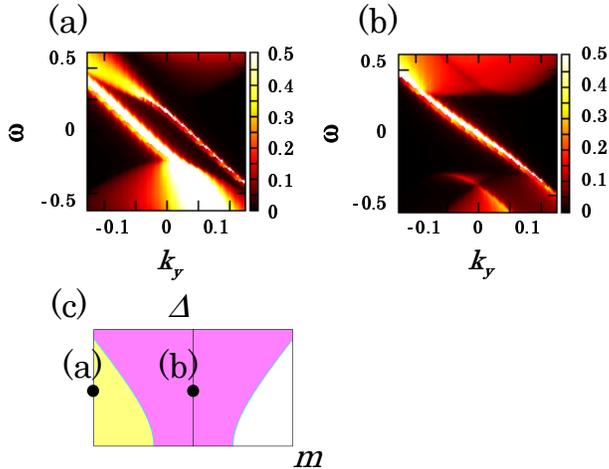}
	\caption{(Color online) Energy spectrum of the edge modes are 
plotted as a function of $k_{y}$ with $\mu$=0.2 and $\Delta$=0.25 
for (a) $m=-0.5$, and (b) $m=0$, respectively. 
(c) Schematic phase diagram for $\mu\neq 0$.
The light yellow region corresponds to ${\cal N}=2$, and the dark red
one to ${\cal N}=1$.
The dots denotes the parameters corresponding to Figs.\ref{fig.9}(a) and (b).}
	\label{fig.9}
\end{center}
\end{figure}

Now, we focus on the edge state for ${\cal N}=2$ with nonzero $\mu$. 
In Fig. \ref{fig.9}(a), the momentum 
resolved LDOS ${\cal N}(k_{y},\omega)$ is plotted 
for $\mu=0.2$, $\Delta=0.25$ and $m=-0.3$.
Since the magnitude of the bulk energy gap $E_{g}$ is expressed as 
\[
E_{g}=\mid \sqrt{m^{2} + (\mu_{B}H)^{2}} - \sqrt{\mu^{2} + \Delta^{2}} \mid
\]
$E_{g}$ is $0.18$ in the present case.
Comparing Fig. \ref{fig.9}(a) with Fig.\ref{fig.3}(a), 
we find that the degenerate two Majorana edge modes realized in ${\cal
N}=2$ phase for $\mu=0$ is lifted off by the doping.  

If we change $m$ from $m=-0.5$ to $m=0$, with fixing $\mu=0.2$ and
$\Delta=0.25$, the corresponding Chern number changes from ${\cal N}=2$
to ${\cal N}=1$. 
In Fig. \ref{fig.9}(b), the momentum 
resolved LDOS ${\cal N}(k_{y},\omega)$ is plotted, 
where $E_{g}$ is given by 0.32.
The chiral Majorana edge mode exists for $\mid \omega \mid < 0.32$. 
By contrast to the edge state of ${\cal N}=2$, that of ${\cal N}=1$ 
is robust against doping.  

\section{Summary}
\label{sec:4}
In this paper, we have studied the edge states of  QAHS coupled with spin-singlet $s$-wave 
superconductor by the proximity effect. We have calculated the energy 
spectrum of the edge states and the resulting local density of states 
for various magnitudes of mass term $m$, pair potential $\Delta$ and 
chemical potential $\mu$. 
The presence or absence of 
chiral Majorana edge modes influence seriously on the 
local density of state. We have clarified that it is possible to 
discriminate the state with 
nonzero Chern number ${\cal N}$ from that with ${\cal N}=0$ 
by observing local density of states by STS. 
Since chiral Majorana edge modes for ${\cal N}=2$ are degenerate each other, 
it is difficult to discriminate ${\cal N}=1$ 
and ${\cal N}=2$ state by simply looking at 
the energy dispersion relation. 
To resolve this problem, we apply the Zeeman magnetic field. 
We find when the direction of the 
magnetic field is parallel to the interface, the degeneracy 
of the present two chiral Majorana edge modes in ${\cal N}=2$ states 
are  lifted off.  
Due to the presence of the bulk edge correspondence, 
the number of the chiral edge modes can be evaluated by the 
winding number defined at $k_{y}=0$. 
%
For ${\cal N}=1$ state, the resulting chiral Majorana 
edge mode are protected   
by the topological property of the bulk Hamiltonian.  
%

\vspace{3ex}
\centerline{\bf Acknowledgment}

We would like to express our sincere thanks to M. Oshikawa for
 giving meaningful advices to us.
And this work was supported in part by the Grant-in Aid for Scientific
Research from MEXT of Japan, "Topological Quantum Phenomena"
No.22103005 (Y.T, M.S.), No.20654030, No.22340096 (Y.T.),  
and No.22540383 (M.S.).

\end{document}